\documentclass{article}

\usepackage{amssymb}
\usepackage{amsmath}
\usepackage[numbers,sort&compress]{natbib}
\usepackage{authblk} 
\usepackage{graphicx}
\usepackage{dcolumn}
\usepackage{bm}
\usepackage{hyperref}
\usepackage{siunitx}
\usepackage{textcomp} 
\usepackage{color} 
\usepackage{bbm} 
\usepackage{soul} 

\numberwithin{equation}{section}

\newcommand{\posner}{Ca\textsubscript{9}(PO\textsubscript{4})\textsubscript{6}} 
\newcommand{\el}[2]{{}\textsuperscript{#1}#2} 
\newcommand{\pyro}{P\textsubscript{2}O\textsubscript{7}\textsuperscript{4-}} 
\newcommand{\ca}{Ca\textsuperscript{2+}} 
\newcommand{\point}[2]{#1\textsubscript{#2}} 
\newcommand{\tc}[1]{$\tau_{\rm #1}$} 

\begin{document}

\title{Posner qubits: spin dynamics of entangled Ca\textsubscript{9}(PO\textsubscript{4})\textsubscript{6} molecules and their role in neural processing}
\author{Thomas C. Player}
\author{P. J. Hore}
\affil{Department of Chemistry, University of Oxford, Oxford OX1 3QZ, UK}
\date{}

\maketitle

\begin{abstract}
It has been suggested that \el{31}{P} nuclear spins in \posner{} molecules could form the basis of a quantum mechanism for neural processing in the brain. A fundamental requirement of this proposal is that spins in different \posner{} molecules can become entangled and remain so for periods (estimated at many hours) that hugely exceed typical \el{31}{P} spin relaxation times. Here, we consider the coherent and incoherent spin dynamics of \posner{} arising from dipolar and scalar spin-spin interactions and derive an upper bound of 37 min on the entanglement lifetime under idealized physiological conditions. We argue that the spin relaxation in \posner{} is likely to be much faster than this estimate.

\end{abstract}

\section{Introduction}

In a recent article entitled ``Quantum Cognition", Matthew Fisher suggested that nuclear spins might act as qubits in neural processing~\cite{fisher_quantum_2015}. He proposed that \el{31}{P} nuclei could be quantum mechanically entangled in networks of `Posner molecules', \posner{}, formed by the enzymatic hydrolysis of pyrophosphate (\pyro{}), and that pairs of Posner molecules might remain entangled for a day ``or possibly much longer"~\cite{weingarten_new_2016}. It was argued that such abnormally long-lived spin coherence could ``modulate quantum correlations between the pairwise binding of multiple Posner molecules, even when the pairs are well separated" in spatially distant neurons, apparently allowing long-range quantum-correlated discharge of \ca{} ions as part of a ``quantum-to-biochemical transduction" mechanism~\cite{fisher_are_2017}.

Long-lived nuclear spin states are well known in NMR spectroscopy~\cite{levitt_singlet_2012, pileio_relaxation_2010, carravetta_beyond_2004, pileio_long-lived_2008, vasos_long-lived_2009, pileio_storage_2010, tayler_singlet_2011, feng_accessing_2012}. Normally involving just two spin-\textonehalf{} nuclei in the same molecule (but see Refs~\cite{stevanato_long-lived_2015,hogben_multiple_2011}), nuclear singlet states have been created that persist for many multiples of the spin-lattice relaxation time $T_1$. In one case, a relaxation time in excess of an hour ($\sim50 T_1$) was measured for a pair of \el{13}{C} spins in a bespoke organic compound~\cite{stevanato_nuclear_2015}. Such states owe their longevity not just to their immunity from dominant spin relaxation pathways but also to careful experimental control of the coherent spin dynamics arising from chemical shifts and $J$-couplings, using field shuttling, spin-locking, and spin-decoupling~\cite{levitt_singlet_2012}. Fisher's proposal, by contrast, concerns a molecule whose spin relaxation is claimed to be extremely slow for \emph{all} nuclear spin states, without the requirement to manipulate the coherent part of the spin Hamiltonian~\cite{fisher_quantum_2015, swift_posner_2018}.

\begin{figure}[tbh]
    \centering
    \includegraphics[width=0.8\linewidth]{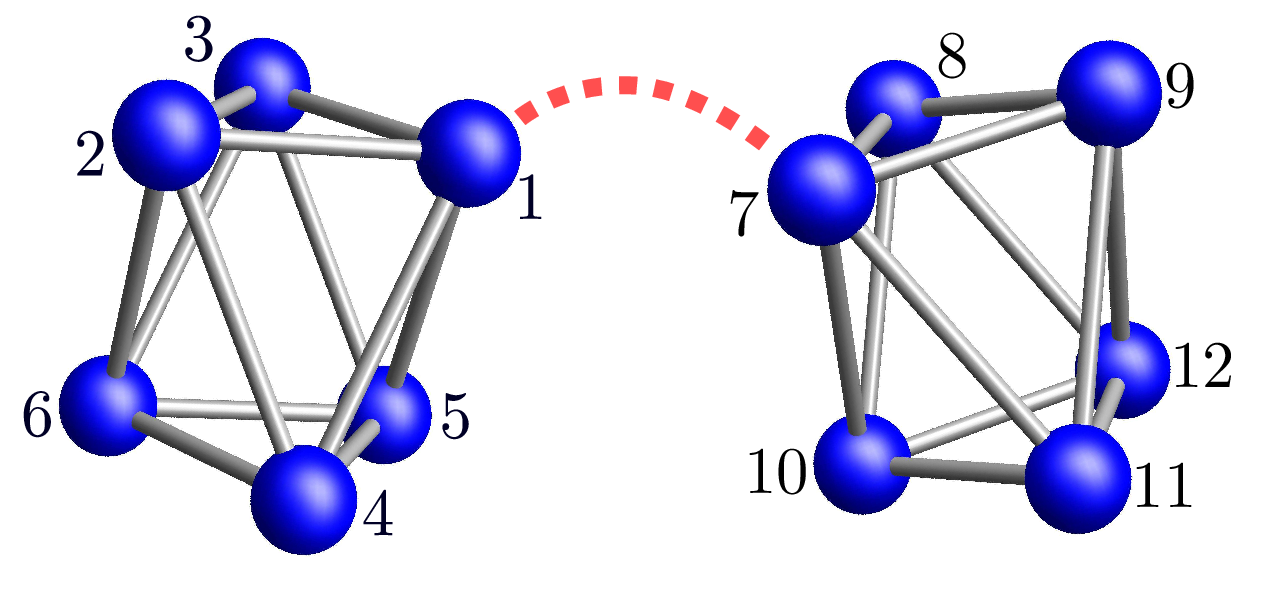}
    \caption{Arrangement of phosphorus atoms in a pair of model Posner molecules \posner{}. The dotted line indicates the two entangled spins in the singlet state $\hat{P}_{1,7}^{\rm S}$ (see section~\ref{methods}). The \point{S}{6} geometry has been emphasized in this representation by exaggeratedly stretching an \point{O}{h} structure along a three-fold rotation axis. Coupling constants~\cite{swift_posner_2018}: nearest neighbours (e.g.\ 1 and 2), $J_{\rm A} = +\SI{0.178}{\hertz}$; next nearest neighbours (e.g.\ 1 and 4), $J_{\rm B} = +\SI{0.145}{\hertz}$; third nearest neigbours (e.g.\ 1 and 6), $J_{\rm C} = -\SI{0.003}{\hertz}$. Atoms in the second Posner molecule are numbered 7--12 in the same order as the first.}
    \label{fig1}
\end{figure}

`Posner's cluster' was originally identified by Betts and Posner as a structural unit in hydroxyapatite~\cite{posner_synthetic_1975}, the main inorganic constituent of bone tissue. It is now thought to be an early intermediate in the nucleation of amorphous calcium phosphate, the precursor of hydroxyapatite~\cite{yin_biological_2003,dey_role_2010, wang_posners_2012,mancardi_detection_2017}. \emph{Ab initio} structure calculations suggest that, \emph{in vacuo}, an isolated Posner's cluster, \posner{}---henceforth referred to as a Posner molecule~\cite{swift_posner_2018}---has eight calcium ions at the vertices, and a ninth in the centre, of a distorted cube, with a phosphate ion near the middle of each cubic face~\cite{treboux_existence_2000}. The phosphorus atoms, positioned at the vertices of an octahedron stretched along one of its three-fold rotation axes, have \point{S}{6} symmetry with a nearest neighbour separation of $\SI{\sim0.5}{\nano\metre}$~\cite{treboux_existence_2000}. Figure~\ref{fig1} shows a representation of the arrangement of the phosphorus atoms in this structure.

Several features of \posner{} suggest slow \el{31}{P} spin relaxation in aqueous solution. (a) \el{31}{P}, the only stable isotope of phosphorus, has spin-\textonehalf{} and therefore no electric quadrupole moment that would be relaxed by locally fluctuating electric field gradients. (b) Of the five stable isotopes of calcium, only \el{43}{Ca}, with a natural abundance of 0.135\%, has non-zero spin. Unlike phosphate ions ($\rm HPO_4^{2-}$ and $\rm H_2PO_4^{-}$), Posner's molecule contains no other magnetic nuclei (e.g.\ \el{1}{H}) that would lead to efficient dipolar relaxation. (c) Although \el{31}{P} relaxation in strong magnetic fields is often dominated by chemical shift anisotropy (CSA), this mechanism will be negligibly slow in the Earth's magnetic field. (d) As a small ($\SI{\sim0.9}{\nano\metre}$ diameter), approximately spherical molecule~\cite{dey_role_2010}, the relatively rapid rotational diffusion of \posner{} would tend to reduce the influence of intermolecular spin-spin interactions. Taken together, these properties conjure up a molecule whose nuclear spins are magnetically isolated from one another and from their surroundings. It was from such considerations that Fisher obtained his original estimate of a 1-day \el{31}{P} relaxation time~\cite{fisher_quantum_2015,weingarten_new_2016}, subsequently revised to $\SI{1.8e6}{\second}\simeq21$ days or ``may be even longer"~\cite{swift_posner_2018}. \el{31}{P} relaxation times of small molecules in mobile liquids are normally a few seconds at most~\cite{gorenstein_phosphorus-31_1984}.

Struck by the many orders of magnitude difference between these numbers, we were prompted to look more closely at the spin dynamics of \posner{} to derive an upper bound on the lifetime of the quantum entanglement of a pair of Posner molecules. In particular, we consider here intramolecular \el{31}{P}--\el{31}{P} dipolar and scalar coupling and the Zeeman interaction of the \el{31}{P} spins with the Earth's magnetic field. Interesting, but beyond the scope of this report, are other aspects of Fisher's proposal, such as the enzymatic reaction that creates the pairs of phosphate ions from which the entangled Posner molecules are assembled, and the chemical/physical nature of the read-out process in which $\rm Ca^{2+}$ ions are simultaneously released from entangled molecules that have been transported into remote neurons~\cite{fisher_quantum_2015,weingarten_new_2016,halpern_quantum_2017,fisher_are_2017}.

\section{Methods}
\label{methods}

We treat \posner{} as a rigid molecule, in which the phosphorus atoms have the distorted octahedral (\point{S}{6}) arrangement shown in figure\ 1. We consider the simplest case of intermolecular entanglement---a pair of identical Posner molecules constructed such that two \el{31}{P} spins, one in each molecule, are initially in the maximally entangled nuclear singlet state, $\hat{P}_{1,7}^{\rm S} = |{\rm S}_{1,7}\rangle\langle{\rm S}_{1,7}|$, where $|{\rm S}_{1,7}\rangle = \left[ |\alpha_1\beta_7\rangle-|\beta_1\alpha_7\rangle\right]/\sqrt{2}$. The other ten spins (labelled 2--6 in one molecule and 8--12 in the other, figure~\ref{fig1}) are assigned maximally mixed initial states. The spin Hamiltonian of this 12-spin system, $\hat{H}(t) = \hat{H}_0 + \hat{H}_1(t)$, has a time-independent part,
\begin{equation}
    \hat{H}_0 = \omega_0 \sum_{k} \hat{S}_{kz} + \sum_{j<k} \sum_{k} 2 \pi J_{jk} \hat{\mathbf{S}}_j \boldsymbol{\cdot} \hat{\mathbf{S}}_k,
    \label{eq1}
\end{equation}
in which $\hat{\mathbf{S}}_k$ and $\hat{S}_{kz}$ are, respectively, the spin angular momentum operator, and its $z$-component, for nucleus $k$. The first term in equation~\eqref{eq1} accounts for the Zeeman interaction with the Earth's magnetic field, taken to have flux density $B_0 = \SI{50}{\micro\tesla}$. The \el{31}{P} Larmor frequency is $|\omega_0/2 \pi|=\gamma_{\rm P} B_0 / 2 \pi = \SI{863}{\hertz}$, where $\gamma_{\rm P} = \SI{10.84e7}{\per\tesla\per\second}$ is the magnetogyric ratio of \el{31}{P}. The six \el{31}{P} spins in each molecule (figure~\ref{fig1}) are magnetically equivalent, with identical chemical shifts. The second term in equation~\eqref{eq1} represents the intramolecular $J$-couplings: values of the three distinct coupling constants, taken from~\cite{swift_posner_2018}, are given in figure~\ref{fig1}. We ignore the tiny natural abundance of \el{43}{Ca} and all intermolecular spin interactions.

$\hat{H}_1(t)$, the incoherent part of the spin Hamiltonian, contains the 15 pairwise dipolar interactions in each molecule, which are time-dependent as a result of rotational diffusion:
\begin{equation}
    \hat{H}_1(t) = - \sqrt{6} \hbar \gamma_{\rm P}^2 \left( \frac{\mu_0}{4\pi}\right) \hat{\hat{R}}_{\rm mol}(t) \sum_{j<k} \sum_{k} \hat{\hat{R}}_{\rm pos}^{(jk)} \hat{T}^{(jk)}_{2,0} / r_{jk}^{3}.
    \label{eq2}
\end{equation}
In equation~\eqref{eq2}, $\mu_0$ is the vacuum permeability, $r_{jk}$ is the distance between spins $j$ and $k$, and $\hat{T}^{(jk)}_{2,0}$ is one of the five second rank irreducible spherical tensor operators for spins $j$ and $k$. $\hat{\hat{R}}_{\rm pos}^{(jk)}$ and $\hat{\hat{R}}_{\rm mol}(t)$ are rotations that define, respectively, the fixed direction of the $j$--$k$ dipolar axis in the molecular frame and the time-dependent orientation of the molecule in the laboratory frame. The atomic coordinates needed to calculate $r_{jk}$ and $\hat{\hat{R}}_{\rm pos}^{(jk)}$ were obtained using density functional theory (see electronic supplementary material).

The coherent spin dynamics driven by $\hat{H}_0$, and the spin relaxation induced by $\hat{H}_1(t)$, together determine the time-dependence of the density operator $\hat{\rho}(t)$ of the 12-spin system:
\begin{equation}
    \frac{{\rm d}\hat{\rho}(t)}{{\rm d}t} = - {\rm i} \hat{\hat{L}} \hat{\rho}(t) \implies \hat{\rho}(t) = {\rm exp}\left(- {\rm i} \hat{\hat{L}} t \right) \hat{\rho}(0)
    \label{eq3}
\end{equation}
where the Liouvillian, $\hat{\hat{L}} = \hat{\hat{H}}_0 + {\rm i} \hat{\hat{\Gamma}}$, contains the commutator superoperator corresponding to $\hat{H}_0$ and the relaxation superoperator $\hat{\hat{\Gamma}}$ (discussed below).

\section{Results}

\subsection{Coherent spin dynamics}

\begin{figure}[tbh]
    \centering
    \includegraphics[width=0.8\linewidth]{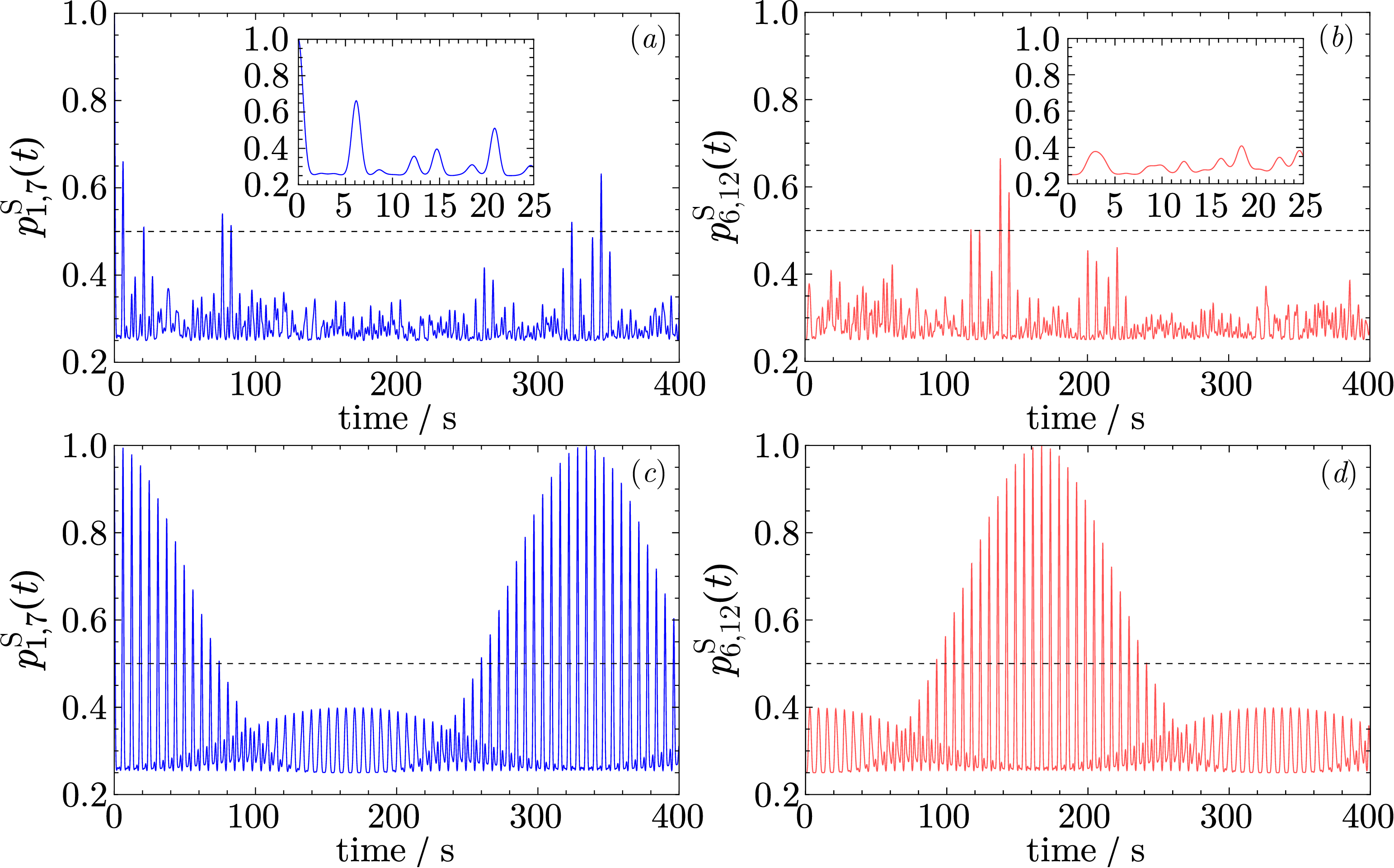}
    \caption{The singlet order $p_{j,k}^{\rm S}(t)$ of spins 1 and 7, (\emph{a}) and (\emph{c}), and spins 6 and 12, (\emph{b}) and (\emph{d}), as a function of time after the formation of a pair of entangled Posner molecules in the initial state $\hat{P}_{1,7}^{\rm S}$. (\emph{a}) and (\emph{b}) are calculated assuming \point{S}{6} symmetry, (\emph{c}) and (\emph{d} with \point{O}{h} symmetry. The dashed horizontal lines indicate $p_{j,k}^{\rm S}=0.5$. The insets in (\emph{a}) and (\emph{b}) show expanded views of the main plots.}
    \label{fig2}
\end{figure}

The initial density operator, $\hat{\rho}(0) = \hat{P}_{1,7}^{\rm S} / 2^{10}$, commutes with the Zeeman interaction but not with the $J$-coupling term in equation~\eqref{eq1}. As a consequence, the singlet order, $p_{1,7}^{\rm S}(t) = {\rm Tr}\left[ \hat{\rho}(t) \hat{P}^{\rm S}_{1,7} \right]$, of the initially entangled pair of spins oscillates at a variety of frequencies determined by the three coupling constants. Ignoring spin relaxation for the moment, $p_{1,7}^{\rm S}(t)$ has the complex time-dependence shown in figure~\ref{fig2}\emph{a}. Within a second, $p_{1,7}^{\rm S}(t)$ drops to 0.25, the value expected for a pair of spins in a maximally mixed state, and only occasionally rises above 0.5 as the various oscillations come partially back into phase. Simultaneously, coherence is transferred to the other ten spins via the $J$-couplings. Although all intermolecular pairs of spins have non-zero singlet order at some point (i.e.\ $p_{j,k}^{\rm S}(t)>0.25$), only $p_{6,12}^{\rm S}(t)$ rises above 0.5 (figure~\ref{fig2}\emph{b}) during the first $\SI{400}{\second}$ (spins 6 and 12 are related to 1 and 7, respectively, by the inversion symmetry, figure~\ref{fig1}).

Figures~\ref{fig2}\emph{c} and \ref{fig2}\emph{d} show $p_{1,7}^{\rm S}(t)$ and $p_{6,12}^{\rm S}(t)$ for a simplified coupling pattern in which the two largest coupling constants have been set equal to their mean, $J_{\rm AB} = \frac{1}{2}\left( J_{\rm A} + J_{\rm B} \right)$. This situation would arise if the four degenerate \point{S}{6} conformations of the molecule were to interconvert under conditions of fast exchange, i.e.\ on a timescale rapid compared to $|J_{\rm A} - J_{\rm B}|^{-1} \simeq \SI{30}{\second}$. As expected, with only two distinct couplings, the time-dependence of the singlet order is somewhat simpler than in figures~\ref{fig2}\emph{a} and \ref{fig2}\emph{b}.

\begin{figure}[tbh]
    \centering
    \includegraphics[width=0.8\linewidth]{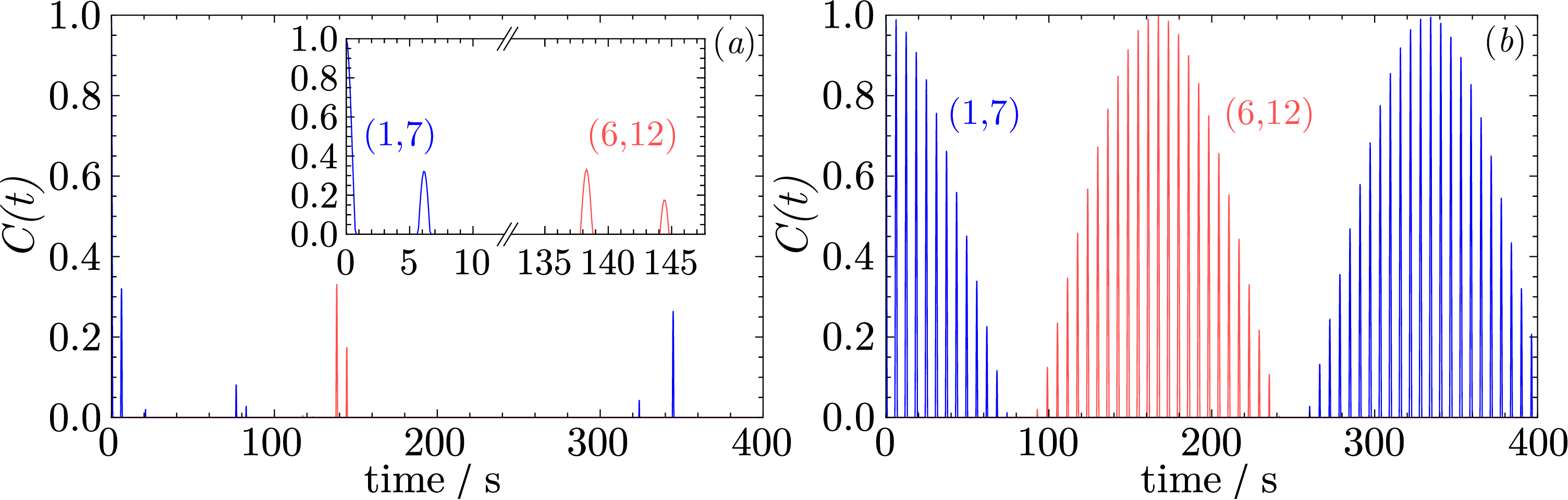}
    \caption{The two-qubit concurrence $C_{j,k}(t)$ of spins 1 and 7 (blue) and spins 6 and 12 (red) as a function of time after the formation of a pair of entangled Posner molecules in the initial state $\hat{P}_{1,7}^{\rm S}$. (\emph{a}) and (\emph{b}) were calculated assuming \point{S}{6} and \point{O}{h} symmetry respectively. The inset in (\emph{a}) shows an expanded view of the main plot.}
    \label{fig3}
\end{figure}

Figure~\ref{fig2} suggests that the coherent spin dynamics may distribute the initial entanglement around the two molecules. The degree of entanglement of spins $j$ and $k$ was quantified by taking the partial trace of the density matrix over the other ten spins and then calculating $C_{j,k}(t)$, the two-qubit concurrence~\cite{wootters_entanglement_1998}. Taking $\hat{\rho}(0) = \hat{P}_{1,7}^{\rm S}/2^{10}$, $C_{j,k}(t)$ was determined for all pairs of spins, for both the full set of couplings $\left\{J_{\rm A},J_{\rm B},J_{\rm C}\right\}$, figure~\ref{fig3}\emph{a}, and the simplified coupling pattern $\left\{J_{\rm AB},J_{\rm AB},J_{\rm C}\right\}$, figure~\ref{fig3}\emph{b}. In both cases, $C_{j,k}(t) = 0$ (i.e.\ no entanglement) for all pairs of spins $(j,k)$ except (1,7) and (6,12) and those pairs are only entangled (i.e.\ $C_{j,k}(t) > 0$) when the corresponding singlet order $p_{j,k}^{\rm S}(t)$ (figure~\ref{fig2}) rises above 0.5. This is the threshold for entanglement of a two-qubit density matrix describing a mixture of $\rm |S\rangle \langle S|$ and $\left( \hat{\mathbbm{1}} - \rm |S\rangle\langle S| \right) /3 $~\cite{hogben_entanglement_2012}. Although for most of the time in figure~\ref{fig3}\emph{a} there is no two-spin entanglement at all, it is possible that intermolecular entanglement involving more than two spins may exist at all times.

\subsection{Spin relaxation}

The most important relaxation pathways for spin-\textonehalf{} nuclei normally arise from rotational modulation of intramolecular dipolar interactions and chemical shift anisotropy. As the latter will be negligibly slow in an Earth-strength magnetic field, we focus on dipolar relaxation, using Redfield theory to obtain $\hat{\hat{\Gamma}}$~\cite{goldman_formal_2001,redfield_theory_1965,slichter_principles_1978}. The validity of this approach is justified in the electronic supplementary material. The two \posner{} molecules are assumed to undergo uncorrelated isotropic Brownian rotational diffusion with a correlation time \tc{c} estimated from the Stokes--Einstein relation, $\tau_{\rm c} = 4 \pi \eta a^3 / 3 k_{\rm B} T$, for a spherical object of radius $a$ in a medium with viscosity $\eta$ at temperature $T$. Taking $a = \SI{0.44}{\nano\metre}$~\cite{dey_role_2010} and $T = \SI{37}{\celsius}$ (approximate physiological temperature), we obtain $\tau_{\rm c} \simeq \SI{58}{\pico\second}$ for a Posner molecule in pure water ($\eta = \SI{6.9e-4}{\kilo\gram\per\meter\per\second}$~\cite{korson_viscosity_1969}). Given that $\tau_{\rm c}^{-1}$ ($\SI{\sim17}{\giga\hertz}$) hugely exceeds the difference between the largest and smallest eigenvalues of $\hat{H}_0$ ($\num{\sim6}|\omega_0| \simeq \SI{33}{\kilo\hertz}$), the extreme narrowing limit can be used to obtain $\hat{\hat{\Gamma}} = - \left\langle \hat{\hat{H}}_1(t) \hat{\hat{H}}_1(t) \right\rangle \tau_{\rm c}$, where $\hat{\hat{H}}_1(t)$ is the commutator superoperator corresponding to $\hat{H}_1(t)$ and the angled brackets indicate the ensemble average necessary to obtain $g(t)$, the correlation function of $\hat{H}_1(t)$ (see electronic supplementary material for details). As is customary in such calculations, we assume $g(t)$ decays exponentially: ${\rm exp}(-t/\tau_{\rm c})$.

The relaxation rate constant of the intermolecular two-spin singlet, $\hat{P}^{\rm S}_{1,7}/2^{10}$, is obtained as
\begin{equation}
    \Gamma_{\rm S} = {\rm Tr}\left[ \hat{P}^{\rm S}_{1,7} \hat{\hat{\Gamma}} \hat{P}^{\rm S}_{1,7} \right]/{\rm Tr}\left[ \hat{P}^{\rm S}_{1,7} \hat{P}^{\rm S}_{1,7} \right]
    \label{eq4}
\end{equation}
(see electronic supplementary material and Ref.~\cite{kuprov_bloch-redfield-wangsness_2007} for details of the calculation).
Equation~\eqref{eq4} gives $\Gamma_{\rm S} / \si{\per\second} = \num{7.86e-6} \tau_{\rm c} / \si{\pico\second}$ and a relaxation time ($\Gamma_{\rm S}^{-1}$) of $\num{\sim37}$ min for Posner molecules dissolved in pure water ($\tau_{\rm c} = \SI{58}{\pico\second}$) at $\SI{37}{\celsius}$. A simple order-of-magnitude check on this result may be obtained as follows. The dipolar spin-lattice and spin-spin relaxation rates for a pair of \el{31}{P} spins with separation $r$ in the extreme narrowing limit are $T_1^{-1} = T_2^{-1} = \frac{3}{2} \left( \hbar \gamma_{\rm P}^2 r^{-3} \mu_0 / 4 \pi \right)^2 \tau_{\rm c}$~\cite{harris_nuclear_1983}. Taking $r \simeq \SI{0.5}{\nano\meter}$ and multiplying this relaxation rate by 10, because each of the entangled spins is relaxed by its five neighbours, we get $\num{\sim1.5e-5} \tau_{\rm c} / \si{\pico\second}$, which is within a factor of two of equation~\eqref{eq4}. For comparison, we have also used equation~\eqref{eq4} to calculate self-relaxation rates for intramolecular singlets. With the atoms numbered as in figure~\ref{fig1}, the three distinct two-spin singlet states, (1,2), (1,4), and (1,6), have self-relaxation times of 51, 43, and 25 min respectively.

\subsection{Long-lived singlet states}

Nuclear singlet states are not relaxed by local magnetic fields whose fluctuations are correlated at the sites of the participating nuclei~\cite{levitt_singlet_2012}. In practice this means that the geometrical arrangement of the spins should have a centre of inversion~\cite{hogben_multiple_2011}. While this is certainly the case for a single Posner molecule, it is clearly not for a pair of Posner molecules undergoing independent rotational diffusion (as shown in figure~\ref{fig1}). We can anticipate therefore that, although there may be one or more long-lived intramolecular singlet states, there is no possibility that singlets comprising spins in more than one molecule could be immune to dipolar (or indeed any other type of) relaxation. The only intramolecular eigenstate of the dipolar relaxation superoperator $\hat{\hat{\Gamma}}$ that has a zero eigenvalue (apart from the identity operator $\hat{\mathbbm{1}}$) is
\begin{equation}
    \left( \frac{1}{2} \hat{\mathbbm{1}} - \hat{P}^{\rm S}_{1,6} \right)\left( \frac{1}{2} \hat{\mathbbm{1}} - \hat{P}^{\rm S}_{2,5} \right)\left( \frac{1}{2} \hat{\mathbbm{1}} - \hat{P}^{\rm S}_{3,4} \right).
    \label{eq5}
\end{equation}
All six phosphorus nuclei are involved in this state in inversion-related, locally singlet pairs. It seems unlikely that this state could arise from the enzymatic hydrolysis of pyrophosphate, which would create singlet pairs of \el{31}{P} spins without the correlation amongst the pairs inherent in equation~\eqref{eq5}. And, being intramolecular, it could not fulfil the proposed quantum cognition function~\cite{fisher_quantum_2015}.

\section{Discussion}

The relaxation time calculated here (37 min) is much smaller than Fisher's most conservative estimate (1 day), obtained by assuming that intermolecular \el{31}{P} dipolar interactions with the protons in rapidly tumbling water molecules would be the dominant relaxation pathway. It is clear from the considerations presented here that intramolecular \el{31}{P}--\el{31}{P} dipolar relaxation is much more efficient. However, 37 min is still a great deal slower than would normally be expected for \el{31}{P}. Given the model employed here, in which relatively weak dipolar interactions are modulated by relatively rapid tumbling, 37 min is not surprising. However, there are several reasons why Posner molecules could relax much more rapidly than this.

(a) The two-spin singlet relaxation rate $\Gamma_{\rm S}$ is proportional to $\tau_{\rm c}$, so anything that hinders the rotation of the molecule will accelerate its \el{31}{P} relaxation. For example: (i) environments more viscous than water; (ii) dimerization (suggested by Fisher to be essential for the read-out of the quantum entanglement~\cite{fisher_quantum_2015}); (iii) formation of larger oligomeric complexes (such as those formed during the nucleation of amorphous calcium phosphate~\cite{posner_synthetic_1975}); and (iv) transient binding to large, slowly moving objects (e.g.\ proteins and membranes).

(b) Other magnetic interactions are unlikely to be negligible. (i) Paramagnetic species such as metal ions and in particular molecular oxygen $\rm O_2$ are effective as relaxation agents by virtue of their large electronic magnetic moments~\cite{tayler_singlet_2011}. (ii) Using \emph{ab initio} calculations with implicit solvation models, Lin and Chiu report two structures of \posner{} that are more stable than the \point{S}{6} form and possess enormous dipole moments (10 D and 31 D)~\cite{lin_structures_2017}. Such structures would be strongly solvated in aqueous solution, giving rise to significant \el{31}{P}--\el{1}{H} dipolar relaxation. (iii) Spin-rotation relaxation (due to the interaction between nuclear spins and the fluctuating magnetic field generated by rapid molecular rotations) may well be significant given the weak dipolar interactions and the absence of CSA relaxation~\cite{pileio_relaxation_2010}.

(c) Lin and Chiu's study suggests that \posner{} may have multiple structures of similar energy~\cite{lin_structures_2017}. If any of them are able to interconvert, our assumption of a rigid, stable molecular framework is unlikely to be valid. (i) Intramolecular rearrangements would modulate both scalar and dipolar \el{31}{P}--\el{31}{P} couplings and so open new relaxation pathways. For instance, degenerate interconversion of \point{S}{6} conformations (see above) occurring in the intermediate exchange regime could destroy the singlet order within seconds. (ii) Intermolecular exchange of phosphate groups, for example:
\begin{equation}
    \rm Ca_9(PO_4)_6 + {}^{*}HPO_4^{2-} \rightleftharpoons Ca_9(PO_4)_5({}^{*}PO_4) + HPO_4^{2-}
    \label{eq6}
\end{equation}
(where * denotes the incoming phosphate), would swap an entangled spin for one with a random spin state and so attenuate the intermolecular spin correlation. Given that \el{31}{P} in free phosphate ($\rm HPO_4^{2-}$ or $\rm H_2PO_4^{-}$, depending on the pH) relaxes in seconds~\cite{gorenstein_phosphorus-31_1984}, exchange reactions of this sort could dramatically accelerate the \el{31}{P} relaxation of \posner{}. (iii) A molecular dynamics study by Mancardi \emph{et al.} found Posner-type clusters with low Ca:P ratios that contained protonated phosphate groups, clusters that shared phosphate groups, and clusters in which $\rm Na^{+}$ is partially substituted for $\rm Ca^{2+}$~\cite{mancardi_detection_2017}. Formation and interconversion of such structures could also contribute to spin relaxation.

Finally, Fisher has recently proposed ``quantum dynamical selection" rules based on the indistinguishability of symmetry-related nuclei in small molecules~\cite{fisher_quantum_2018}. Two consequences are that homolytic bond-breaking of ortho-H\textsubscript{2} should be symmetry-forbidden and that entanglements of nuclear spin states with molecular rotations (e.g.\ in Posner's molecule) could persist ``for exponentially long times". Such arguments seem to overlook the inevitability of nuclear spin relaxation. Wigner showed in 1933 that ortho- and para-H\textsubscript{2} are interconverted when the hydrogen nuclei experience different local magnetic fields~\cite{wigner_uber_1933}. Subsequently Curl \emph{et al}.\ established that dipolar and spin-rotation interactions can rapidly equilibrate nuclear spin isomers in larger molecules~\cite{curl_nuclear_1967}. These interactions couple states of different nuclear and rotational symmetry because they have components that are antisymmetric with respect to exchange of both the positions and the spins of symmetry-related nuclei~\cite{petzinger_para-_1973,steiner_magnetic_1989}. Fast interconversion of spin isomers would render any selection rules immaterial.

\section{Conclusions}

There are three main conclusions. (a) The $J$-couplings of the \el{31}{P} spins in a non-interacting pair of \posner{} molecules drive coherent spin dynamics in which the initial two-spin order is spread around the 12 spins and varies from one second to the next. Whether this is consistent with Fisher's quantum cognition mechanism will probably depend critically on the details of the chemical process that `reads out' the information carried by the spins in the two distant molecules. (b) There are no long-lived intermolecular singlets that are immune to either the coherent or incoherent parts of the spin Hamiltonian. (c) A strict upper limit on the lifetime of the initial intermolecular entanglement of 37 min has been obtained. Various factors have the potential to accelerate this relaxation; we suspect a lifetime of a few seconds is more likely than half an hour. Indeed, the only way we can envisage a molecule with \el{31}{P} relaxation as slow as 1 min under physiological conditions would be a two-spin \emph{intramolecular} nuclear singlet (which would be useless as a source of ``spooky action at a distance"~\cite{einstein_can_1935}).

\vspace{3em}

\emph{Data accessibility.} This article has no additional data.

\emph{Author's contributions.} P.J.H.\ conceived the study. T.C.P.\ performed the study. Both authors discussed the results and wrote the manuscript.

\emph{Competing interests.} We declare we have no competing interests.

\emph{Funding.} This work was supported in part by the European Research Council under the European Union's 7\textsuperscript{th} Framework Programme, FP7/2007--2013/ERC Grant 340451.

\emph{Acknowledgements.} We are grateful to Matthew Fisher, Malcolm Levitt, Jonathan Jones, and Daniel Kattnig for helpful comments on the manuscript and to Daniel Kattnig for DFT calculations of atomic coordinates.

\bibliography{posner_bibliography}
\bibliographystyle{vancouver}

\end{document}